\documentclass{aa}
\usepackage[varg]{txfonts}
\usepackage{wasysym}
\usepackage{graphicx}
\usepackage{gensymb}
\usepackage{subcaption}
\usepackage{arydshln}
\usepackage{rotating}
\usepackage{appendix}

\usepackage{txfonts}
\begin{document}

   \title{Characteristics and evolution of sheath and leading edge structures of interplanetary coronal mass ejections in the inner heliosphere based on Helios and Parker Solar Probe observations}

  \titlerunning{Characteristics and evolution of sheath and leading edge structures of interplanetary coronal mass ejections in the inner heliosphere}

   \author{M. Temmer
          \inst{1}
          \and
          V. Bothmer
          \inst{2}
          }

   \institute{Institute of Physics, University of Graz,
              Universit\"atsplatz 5, A-8010 Graz\\
              \email{manuela.temmer@uni-graz.at}
         \and
            Institute for Astrophysics and Geophysics, University of Göttingen, Friedrich-Hund-Pl.1, 37077 Göttingen, Germany\\
             \email{volker.bothmer@uni-goettingen.de}
             }

   \date{Received Dec xx, 2021; accepted xx}

 
  \abstract
   {We investigate the plasma and magnetic field characteristics of the upstream regions of interplanetary coronal mass ejections (ICMEs) and their evolution as function of distance to the Sun in the inner heliosphere. Results are related both to the development of interplanetary shocks, sheath regions and compressed solar wind plasma ahead of the magnetic ejecta.}
   {From a sample of 45 ICMEs observed by Helios 1/2 and Parker Solar Probe we identify four main density structures, namely shock, sheath, leading edge, and magnetic ejecta (ME) itself. We derive separately for each structure averaged parameter values, such as proton particle density, plasma-beta, temperature, magnetic field strength, proton bulk speed, and duration and place the results into comparison with the upstream solar wind in order to investigate the interrelation between the different density structures. }
   {For the statistical investigation, we use plasma and magnetic field measurements from 40 well-observed Helios 1/2 events during 1974--1981 on the basis of the ICME list compiled by \cite{bothmer98}. Helios data cover the distance range from 0.3--1\,au. For comparison, we add a sample of five ICMEs observed with Parker Solar Probe 2019--2021 over the distance range 0.32--0.75\,au.}
   {It is found that the sheath structure presumably consists of compressed plasma as a consequence of the turbulent solar wind material following the shock and lies ahead of a region of compressed ambient solar wind. The region of compressed solar wind plasma is typically found directly in front of the magnetic driver and seems to match the bright leading edge commonly observed in remote sensing observations of CMEs. From the statistically derived density evolution over distance, we find the CME sheath to become denser than the ambient solar wind at about 0.06\,au. Between 0.09--0.28\,au the sheath structure density starts to dominate over the density within the ME. The ME density seems to fall below the ambient solar wind density over 0.45--1.07\,au. Besides the well-known expansion of the ME, the sheath size shows a weak positive correlation with distance, while the leading edge seems not to expands with distance from the Sun. We further find a moderate anti-correlation between sheath density and local solar wind plasma speed upstream of the ICME shock. An empirical relation is derived connecting the ambient solar wind speed with sheath and leading edge density. Constraints to that results are given.}
   {The average starting distance for actual sheath formation could be as close as 0.06\,au. The early strong ME expansion quickly ceases with distance from the Sun and might lead to a dominance in the sheath density between 0.09--0.28\,au. The leading edge can be understood as a separate structure of compressed ambient solar wind directly ahead of the ME and is likely the bright leading edge of CMEs often seen in coronagraph images. The results allow for better interpretation of ICME evolution and possibly the observed mass increase due to enlargement of the sheath material. The empirical relation between sheath and leading edge density and ambient solar wind speed can be used for more detailed modeling of ICME evolution in the inner heliosphere.}

   \keywords{solar coronal mass ejections -- solar wind --heliosphere -- observations -- solar terrestrial relations
               }

   \maketitle

\section{Introduction}

\begin{figure}
    \centering
    \includegraphics[width=1.\hsize]{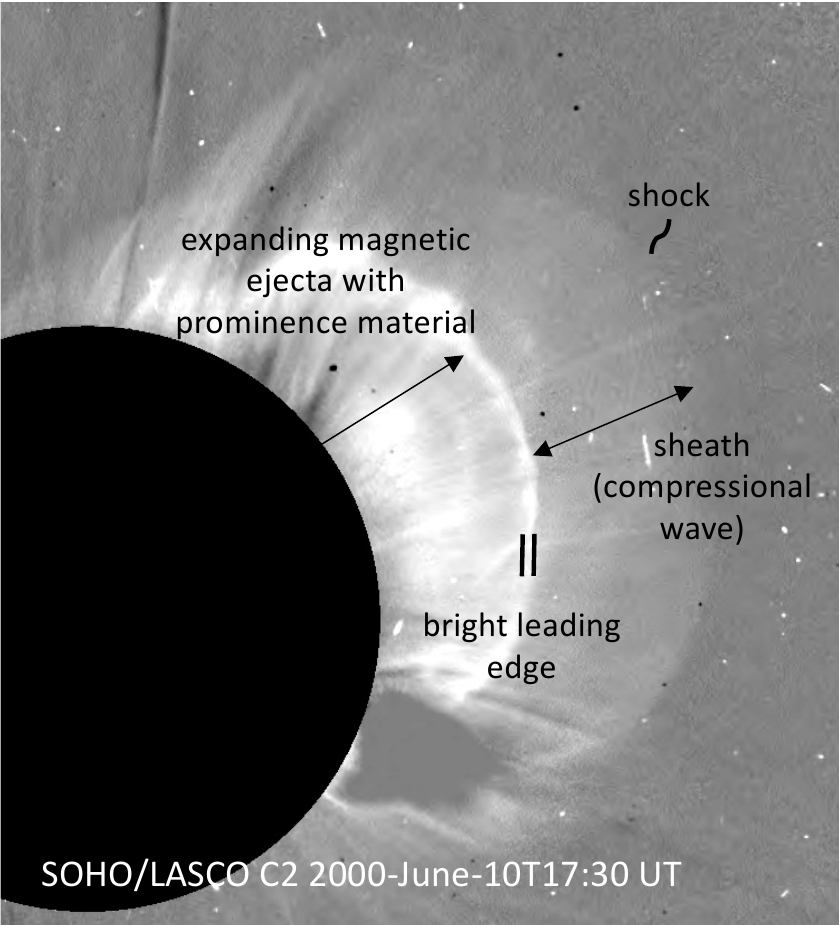}
    \caption{Schematic overview of the different brightness features identified in a coronagraph image of SOHO/LASCO/C2 that occurred on 17 June 2000. See also \cite{vourlidas13}.}
    \label{fig:1a}
\end{figure}

Coronal mass ejections (CMEs) are magnetized plasma volumes that are sporadically ejected from the solar corona and propagate through interplanetary space. They typically consist of several structures as identified in in-situ plasma and magnetic field measurements, but less well in image data showing CME features of different brightness. These are affected by plane-of-sky projection effects and constraints of spatial and time resolution as well as brightness limitations in white-light images of CME events \citep[see e.g.,][]{vourlidas13,kwon17}. In white-light coronagraph data, fast CME events often reveal blurry fronts, which show the propagating shock wave ahead of the CME bright front, the so-called bright leading edge. Behind the bright leading edge typically a dark cavity occurs likely due to the lower density of the expanding magnetic structure, commonly interpreted as a magnetic flux rope. Sometimes in the trailing edge or also within this cavity an intensity enhancement is observed, the so-called bright core, which is typically interpreted as ejected prominence/filament material dragged out with the expanding flux rope structure (see Figure~\ref{fig:1a}). Using in-situ measurements at 1~au, \cite{kilpua13} aimed to relate the in-situ to the white-light structures and classified features measured for interplanetary CMEs (ICMEs) into shock-sheath, flux rope disturbed front regions, magnetic flux rope body and a trailing rear flux rope part. Recent studies followed this approach and treated the frontal region as separate segment with the characteristics of enhanced density and flow speed \cite[e.g.,][]{nieves18,maricic20}. However, there is considerable discussion about the proper interpretation of solar wind characteristics observed in-situ ahead of the ICME magnetic ejecta part and also a lack of statistical studies investigating these structures as function with distance from the Sun. We use the term ICME to predominantly describe the magnetic ejecta, with the structures shock, sheath, and leading edge as a consequence of and being driven by the ICME.  

Strong shock waves and fast magnetic ejecta with a long-lasting negative B$_z$ component have been identified to be the main driver of the strongest space weather effects and are therefore most often the focus of ICME studies. The generation process and development of the magnetized plasma structures and upstream regions ahead of the magnetic ejecta, however, are less well investigated but also play a crucial role in triggering geomagnetic storms \citep[e.g.,][]{gosling87,mccomas88,mccomas89,zurbuchen06,lugaz16}. Their characteristics are not well understood as well as their interplanetary evolution and relationship to the overall ICME mass estimations \citep[mass accretion, see e.g.,][]{deforest13,temmer21}. 

The early strong lateral expansion behavior of fast CMEs may be interpreted as combined bow/piston, driving shock waves in the corona \citep[e.g.,][]{temmer09,patsourakos09}. For strong CME events that move considerably faster than their coronal and solar wind environment, the wave propagating ahead of the strongly expanding driver, i.e., the magnetic ejecta, may steepen into a shock \citep[e.g.,][]{zic08}. Observational studies showed that the CME mass increases found close to the Sun (up to 0.1\,au) based on the analysis of remote sensing data, is most probably related to outflows from the Sun \citep{bein13,howard18}. These outflows form well in temporal association with coronal dimming regions at the footpoints of a CME, and could be directly related to the early CME mass evolution \citep{temmer17,dissauer19}. Composition measurements revealed that the sheath region between shock and CME consists of ambient solar wind material, while the CME is clearly representing coronal material \citep[see e.g.,][]{kilpua17}. However, \cite{deforest13} found from a case study combining EUV, white-light and in-situ data, that the sheath consists of compressed solar wind AND coronal material. A similar conclusion is given in a more recent case study by \cite{lugaz20} investigating an ICME from Mercury to Earth. The expansion of the ICME in IP space depends on the pressure balance and interaction processes with the ambient solar wind and may occur as strong or weak, in a self-similar or non-self-similar manner \citep[e.g.,][]{vrsnak19,gopalswamy20,luhmann20}. As the solar wind is unable to fully flow around that growing structure, there might be a tendency of plasma pile up at the ICME apex \citep{siscoe08}. A study by \cite{temmer21} derived at 1\,au a strong linear anti-correlation between the sheath density and the solar wind speed measured 24 hours before the shock-sheath arrival. How the sheath develops in IP space seems also depend on the magnetic driver part, as found by \cite{masias16,salman21} who used large samples of in-situ measured ICMEs from single and multiple spacecraft close to 1\,au. The variation of the magnetic field inside sheaths is another important topic, with respect to particle acceleration. It is found that turbulence and compression dominates and that substructures, such as small-scale flux ropes, may be located inside \citep[][]{kilpua20}. \cite{masias16} concluded a ``progressive sheath-MC transition'' that might be related to continuous magnetic reconnection happening between sheath and the magnetic ejecta part.

\begin{figure*}
    \centering
    \includegraphics[width=0.9\hsize]{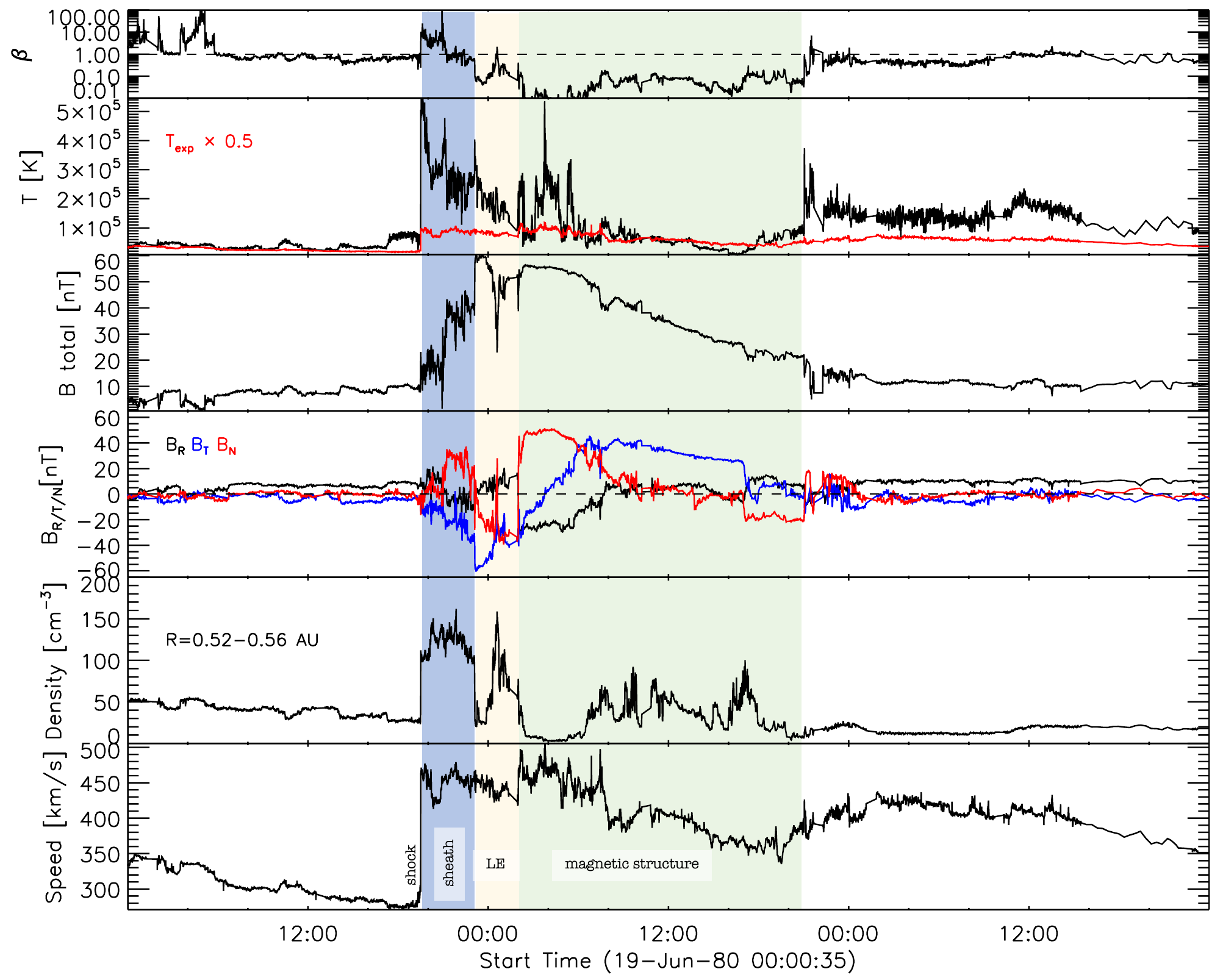}
    \caption{This overview plot shows the plasma and magnetic field data for a 
    unique ICME event observed in June 1980 measured with Helios 1 at 0.53\,au \citep[DOY171--173; see also][]{burlaga82,bothmer98}. From top to bottom we present the parameters plasma-beta, proton temperature \cite[in red we give the expected solar wind temperature $\times$ 0.5, as described in][]{richardson95}, total magnetic field strength, vector components of the magnetic field, proton density and proton bulk speed. The four identified density structures are marked, with the shock arrival discontinuity followed by the sheath region (blue shaded area) ahead of the density enhancement of the so-called LE (yellow shaded area) ahead of the ME body (green shaded area). It should be noted that the green labeled interval matches well with a bi-directional flow of suprathermal electrons covering the period June 20, 02--21\,UT \cite[see Figure 6 in][]{bothmer99}.}
    \label{fig:1}
\end{figure*}

\begin{figure*}
    \centering
    \includegraphics[width=0.85\hsize]{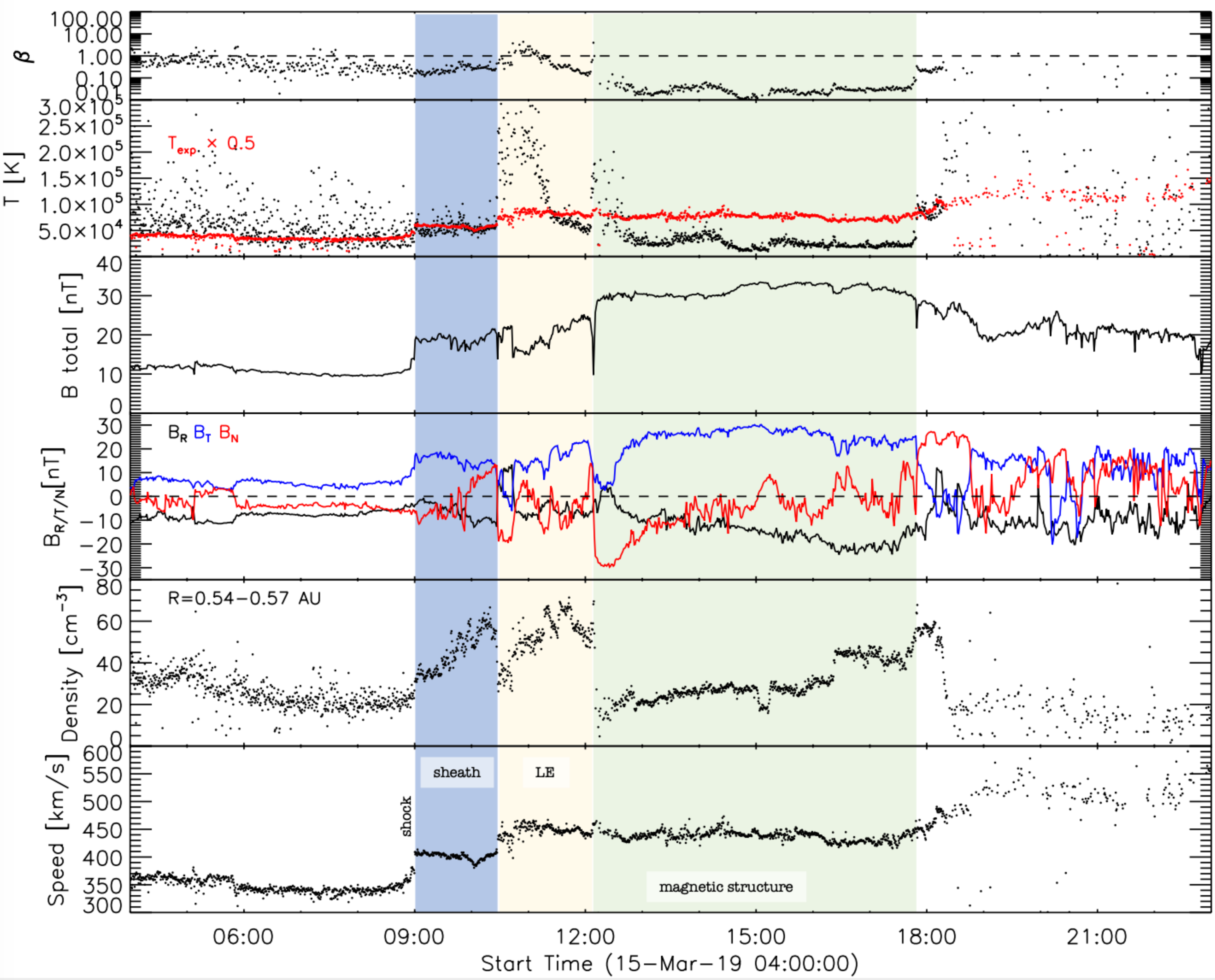}
    \caption{This overview plot shows the plasma and magnetic field data for an ICME observed with PSP in March 2019 (DOY74). Panel parameters and identified structures are the same as shown in Figure~\ref{fig:1}. The plasma data are partially noisy. The time interval of the ICME has been identified by \cite{lario20}.}
    \label{fig:1b}
\end{figure*}

In the current study we focus on different density structures that can be identified from ICME in-situ measurements. We aim to better understand the physical processes between shock and ICME body structure, hence ME region. We will show that, comparable to many white-light observations of CME events, typically fast CMEs reveal two density enhancements that can be related to the shock-sheath and leading edge regions ahead of the magnetic ejecta commonly identifiable by the strong decrease of the plasma-beta. Further we aim to provide more details about the physical characteristics of these different ICME driven structures in IP space with distance from the Sun and address the question of a possible ``starting'' distance of sheath formation and also on how local solar wind conditions might be related to the sheath properties. For this purpose we investigate a Helios data set of 40 well-observed ICMEs over the distance range 0.3--0.9\,au. For comparison, we study the structure densities for a set of five observed ICME events between 0.32--0.75\,au by the recently launched Parker Solar Probe (PSP) mission.

In Section \ref{sec2} we describe the data and methods and the results in Sections \ref{sec3} and \ref{sec4}. Discussion and conclusion of the outcomes are presented in Section \ref{sec5}.

\section{Data and Methods}\label{sec2}
 
We investigate in-situ plasma and magnetic field measurements from two missions covering a wide distance range in the inner heliosphere. Our main focus lies on the Helios 1/2 spacecraft with a perihelion around 0.3~au \citep{rosenbauer77}. From Helios 1/2, using hourly averaged plasma and magnetic field data, we identify for 40 well-observed events covering the time range 1975--1981 the ICME structures sheath, leading edge (LE) and magnetic ejecta (ME). The identification of these ICMEs is based on the list given in \cite{bothmer98}. An example of the ICME event in June 1980 from Helios 1 at 0.53~au \citep[DOY171--173; see also][]{burlaga82,bothmer98} is shown in Figure~\ref{fig:1}. In the Appendix we show the in-situ measurements and identified structures for another sample of Helios ICME events (Figure~\ref{fig:helios_app}) and provide a table with event dates and distances (Table~\ref{tab:helios}).

For comparison, we cross-check the results with in-situ measurements from the currently operating PSP mission using data from the SWEAP and FIELDS investigations \citep[][]{bale16,kasper16}. PSP will encounter the Sun in the near future as close as up to 0.046\,au \citep{fox16}. Using the regularly updated list of ICMEs measured by PSP as compiled by Bothmer and Chifu (see \url{http://cgauss-psp.astro.physik.uni-goettingen.de/pro_work.php}), we investigate a set of five ICME events observed in 2019--2021 over the distances 0.32--0.75\,au. Compared to Helios mission time, the solar activity is weaker, hence, we currently miss events with strong shock and sheath regions. From daily plots of 1 minute resolution plasma and magnetic field data we identify the different ICME structures and derive the average density for each of it. An example of the ICME event in March 2019 from PSP at 0.55\,au \citep[DOY74; see also][]{lario20} is shown in Figure~\ref{fig:1b}. Table~\ref{tab:psp} gives PSP event dates, distances and references for each event. In the Appendix we show the in-situ measurements and identified structures for the other PSP ICME events (Figure~\ref{fig:psp_app}). 

We first investigate the plasma and magnetic field data to identify the different structures ahead of the ICME (see Figures~\ref{fig:1} and \ref{fig:1b}). After the shock discontinuity, we identify the sheath structure by its high density, temperature, plasma-beta larger than one, and strongly fluctuating magnetic field. The LE is identified by a moderate plasma-beta, increased magnetic field and density, and discontinuities before and after, marking the end of the sheath and start of the ME, respectively \citep[see also][]{wimmer06,kilpua13,maricic20}. The ME structure itself reveals very low plasma-beta, low temperature together with a smooth and rotating magnetic field vector and decreasing speed profile. Producing such plots from Helios 1/2 plasma and magnetic field measurements and using the criteria described above, we manually identify the start and end time of the (shock-)sheath region, the LE and the ME part. For each identified structure, we extracted the density, bulk speed, total magnetic field strength, temperature and duration. In addition, we investigate for speed and density the up- and downstream conditions and calculate the average of these parameters over 3 hours before and after the shock front arrival. Further parameters are derived and statistically related to each other by calculating the Spearman correlation coefficient on a 90\% confidence level. We give a correlation matrix of the results in the Appendix (Figure~\ref{fig:rplot}).

\section{Results}\label{sec3}

The main focus lies on the large statistical ICME events from Helios 1/2 data. For inspecting the conditions in the up- and downstream regions, we first derive the Alfvén Mach number and density compression ratio for each event. The Alfvén Mach $M_A$ is calculated as $M_A={u}/{(B/\sqrt{\rho \mu_0})}$ where $u$ is the upstream solar wind speed, $B$ is the average magnetic field strength of the sheath region, $\rho$ is the sheath mass density (assuming 4\% helium and 96\% protons) and $\mu_0$ is the permeability of vacuum. From that we derive for $M_A$ an average value and standard deviation of 5.54$\pm$2.37. For the density jump, given as ratio over down- and upstream region $\rho_d/\rho_u$, we derive 2.97$\pm$1.59, and we find a value of 3.05$\pm$1.62 when applying the average density over the entire sheath duration instead of the 3 hours averaged density after the shock arrival. While the Alfvén Mach number shows a weak dependency over distance, the derived values for the density jump are randomly distributed over distance. The correlation matrix in the Appendix reveals, that the Alfvén Mach number is, as expected, positively correlated to the magnetic field strength and fluctuations in the magnetic field of the ICME structures sheath and LE. The density fluctuation in the LE is moderately and in the ME weakly correlated to the Mach number, however, there is no significant correlation found to the density parameter of the sheath structure.

\begin{figure}
    \centering
    \includegraphics[width=\hsize]{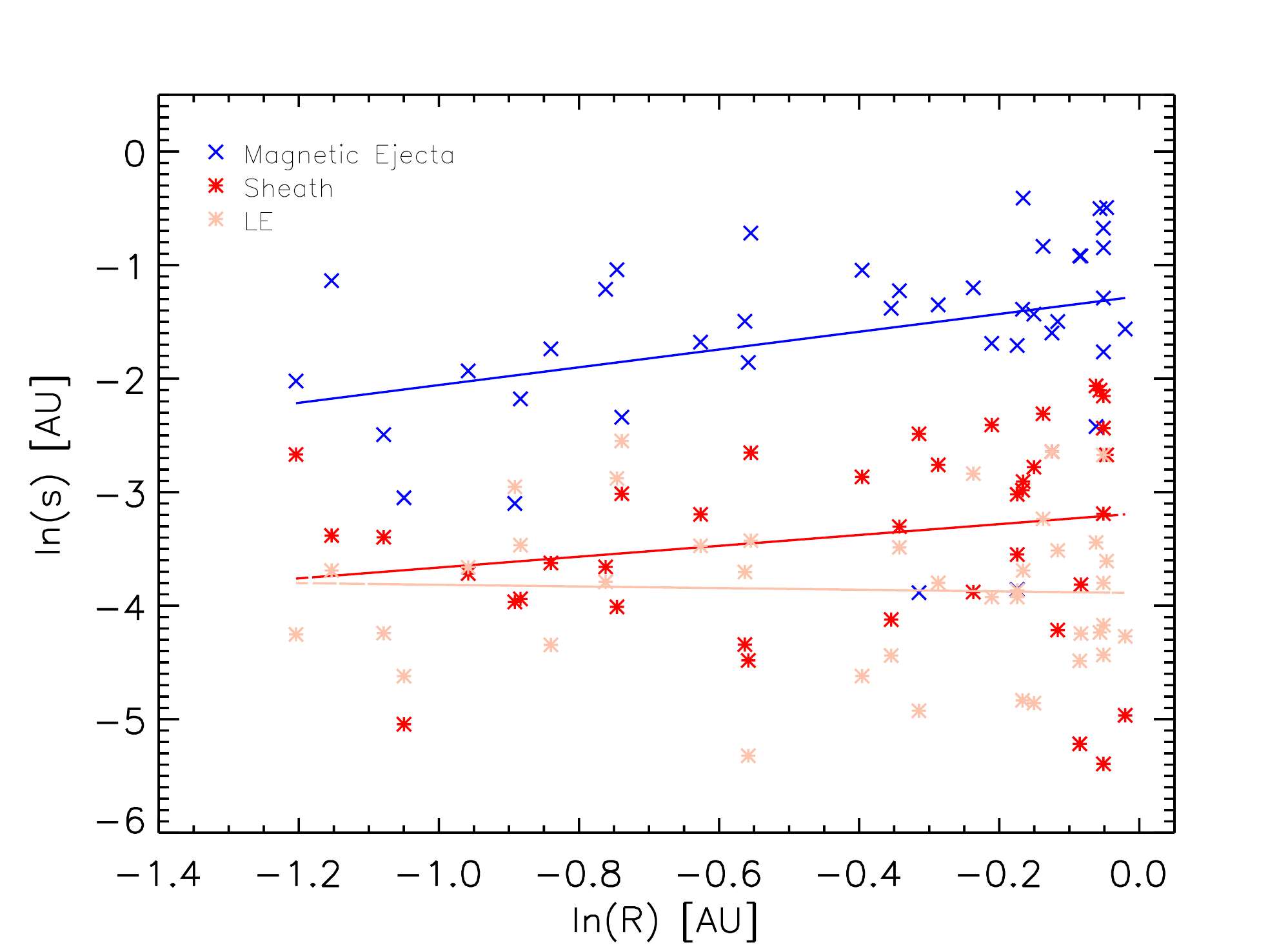}
    \caption{Double logarithmic plot showing the radial sizes, $s$, of the different ICME structures (see legend) versus solar distance, $R$. The linear regression lines for each structure are obtained with: $s_{\rm ME}(R)=0.27 \times R^{0.78}$; $s_{\rm sheath}(R)=0.04 \times R^{0.48}$; $s_{\rm LE}(R)=0.02 \times R^{-0.07}$.}
    \label{fig:14bs}
\end{figure}

Figure~\ref{fig:14bs} shows the radial size of each structure as function of distance. The radial size is simply determined by multiplying the duration of the structure with its average speed. A linear fit is provided for the data points from each structure in order to study its variation in size over distance with $R$ in au. We derive for the ME a linear trend with $s_{\rm ME}(R)=0.27 \times R^{0.78}$, for the sheath structure $s_{\rm sheath}(R)=0.04 \times R^{0.48}$ and for the LE $s_{\rm LE}(R)=0.02 \times R^{-0.07}$. Not surprising, the ME shows the typical expansion behavior in the inner heliosphere which causes a strong increase in radial size with distance. We find the sheath structure enlarging with distance from the Sun, while the LE structure is actually not showing a clear increase in size. Calculating the Spearman correlation coefficient on a 90\% confidence level, the ME shows a moderate correlation with distance with $cc$=0.44, the sheath a weak one with $cc$=0.26, and no significant relation is found for the LE (see Appendix). This finding could be interpreted that the LE is rather a distinct feature from the ambient solar wind flow and shows low evolutionary aspects. We find no obvious relation between Mach number and sheath or LE duration. 

\begin{figure}
    \centering
    \includegraphics[width=\hsize]{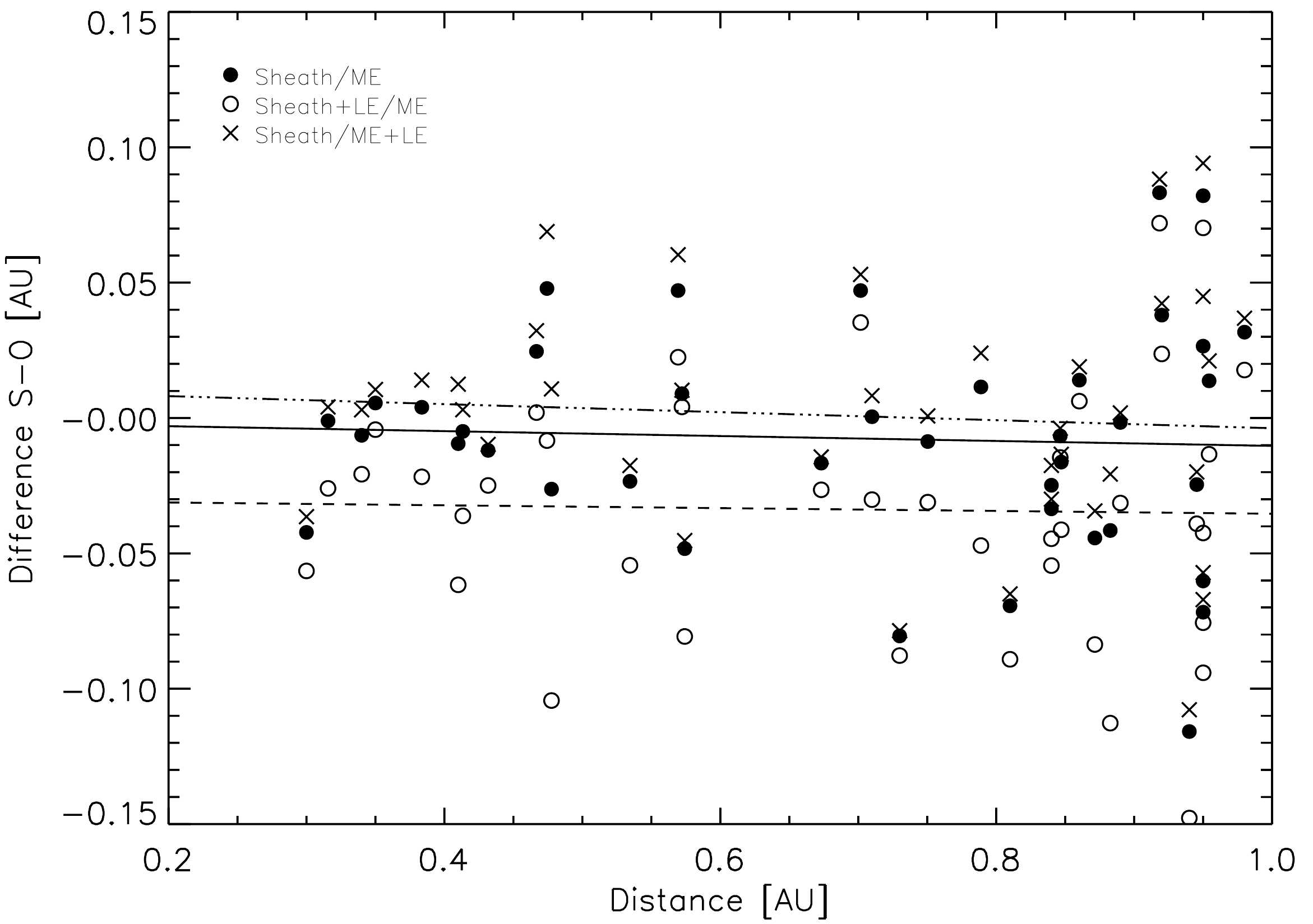}
    \caption{Difference in shock standoff distance between values calculated using the empirical relation for spherical objects (S) given in \cite{seiff62} and the standoff distance as derived from observations (O) of the ICME structures sheath as well as sheath+LE (see symbols in the legend). For the Seiff relation we used different obstacle sizes, ME half-width (size as given in Fig.~\ref{fig:14bs} divided by 2) and ME half-width plus the LE size (ME+LE). The dashed-dotted, solid and dashed lines give linear fits to the differences found for sheath/ME+LE, sheath/ME, and sheath+LE/ME.}
    \label{fig:standoff}
\end{figure}

We perform a simple exercise to find out more about the physical nature of the structures sheath and LE. Assuming that the sheath size is related to the shock standoff distance, $\Delta$, we use the empirical relation for a spherical obstacle given in \cite{seiff62} to calculate the standoff distance. The empirical relation is defined as $\Delta/D=0.78 \rho_u/\rho_d$, where $D$ represents the radius of the obstacle and $\rho_u/\rho_d$ the ratio between upstream ambient solar wind density and sheath downstream density, respectively. For the Seiff relation we used different obstacle sizes, $D$, namely on the one hand the ME half-width (size as given in Fig.~\ref{fig:14bs} divided by 2) and on the other hand the ME half-width plus the LE size. The calculated values are compared to the sheath size as well as to the sum of the sheath and LE size. Figure~\ref{fig:standoff} shows the difference between calculated and observed shock standoff distances as function of distance together with linear fits. We derive for distances beyond 0.45\,au a clear increase in the scatter and the linear fits show a slight negative slope, that can be interpreted as deviation from the spherical shape of the ME as it expands \citep[see also][]{savani12}. Largest deviations from zero are found when comparing the Seiff values, using the ME half-width as obstacle size, to the sheath plus LE size. Best match is found when the Seiff values are compared to the observed sheath size. The structures sheath and LE seem to have a different nature.

\begin{figure*}
    \centering
    \includegraphics[width=0.8\hsize]{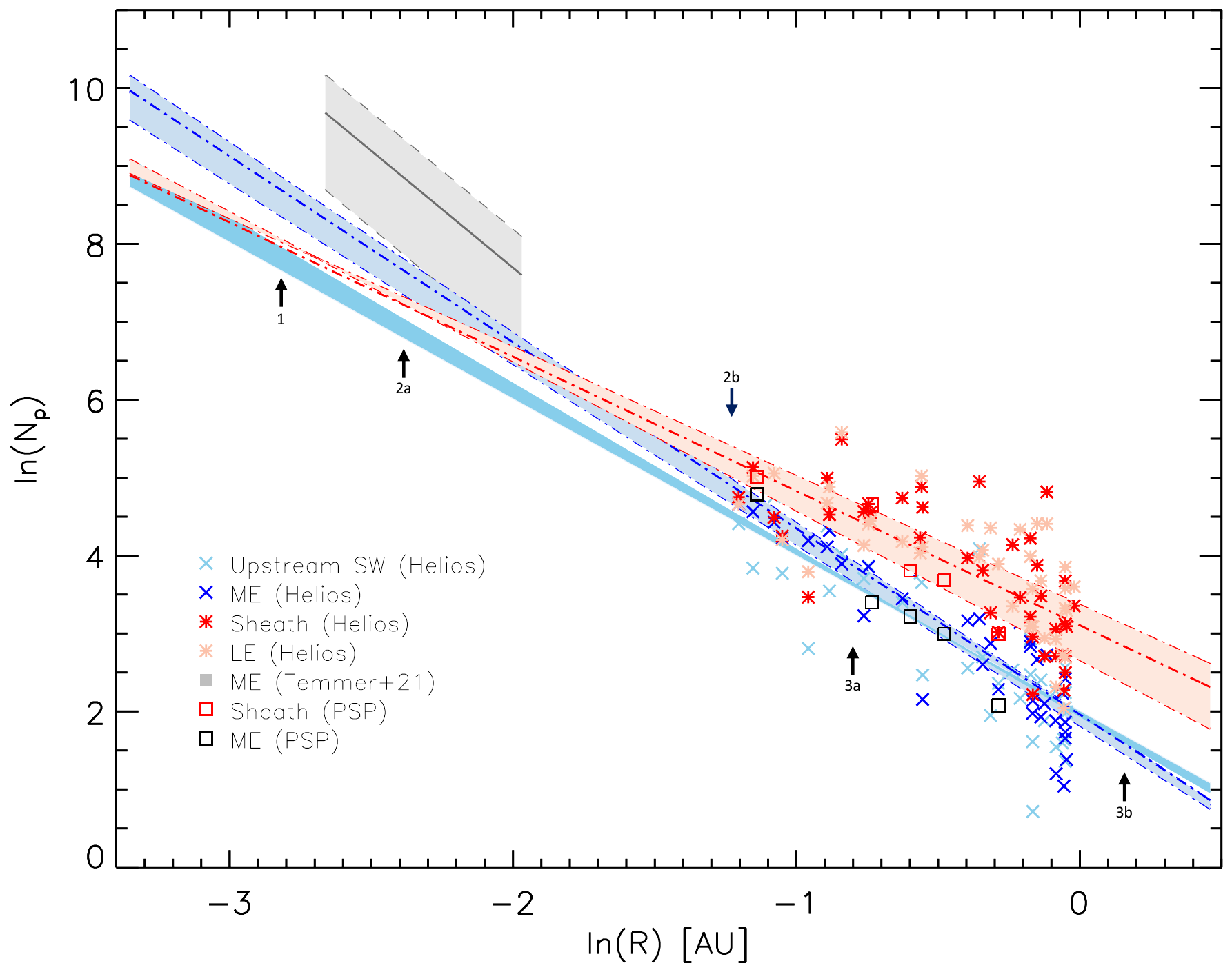}
    \caption{Double logarithmic plot showing the average proton density, $N_p$, of the different ICME structures and upstream solar wind (SW) versus solar distance, $R$, in au. The linear regression lines for each structure from our study are obtained with: $N_{\rm p,SW}(R)=7.0 \times R^{-2.1}$; $N_{\rm p, ME}(R)=7.1 \times R^{-2.4}$; $N_{\rm p,sheath}(R)=22.3 \times R^{-1.7}$; $N_{\rm p,LE}(R)=26.6 \times R^{-1.5}$. The uncertainty ranges for the ME density are derived from previous Helios results from \cite{liu05, wang05,leitner07}. For the SW density estimate we added model results given in \cite{scolini21}. For the sheath density we use as upper and lower uncertainty the standard deviation as derived in this study. Extrapolating the Helios based fits for ME and sheath density, three intersection sectors are derived (1, 2a--2b, 3a--3b which are marked by black arrows). 1: between sheath density and upstream solar wind density at about 0.06\,au - which can be interpreted as average starting distance for sheath formation; 2a--2b: between sheath density and ME density ranging from 0.09--0.28\,au; 3a--3b: between ME density and the upstream solar wind density ranging from 0.45--1.18\,au. Complementary we show PSP sheath and ME density measurements for a set of five events supporting the general trend derived from Helios observations. In addition, we provide results for the mean calculated ME density (solid gray line with upper and lower limit, i.e., standard deviation, marked as light gray dashed lines) over 0.07--0.14\,au based on indirect ME density measurements derived by \cite{temmer21}.}
    \label{fig:15bs:ext}
\end{figure*}

Figure~\ref{fig:15bs:ext} shows the density evolution of each structure and the ambient solar wind in a double logarithmic plot covering the distance range 0.03--1.6\,au. The linear regression line for each structure is obtained with the values of $N_{\rm p,SW}(R)=7.0 \times R^{-2.1}$ for the upstream solar wind, $N_{\rm p,ME}(R)=7.1 \times R^{-2.4}$ for the ME, $N_{\rm p,sheath}(R)=22.3 \times R^{-1.7}$ for the sheath, and $N_{\rm p,LE}(R)=26.6 \times R^{-1.5}$ for the LE. For obtaining additional uncertainty estimates we use Helios results from previous studies and add upper and lower boundaries for the ME density \citep[][]{liu05,wang05,leitner07}. For an uncertainty estimate of the density evolution of the ambient solar wind, we apply results from observations \citep[][]{venzmer18} as well as recent model results from e.g., \cite{scolini21}, and use the standard deviation for the uncertainty estimate of the sheath density. Inspecting the linear fits for ME and sheath density, we obtain three intersection sectors. The first (1) close to the Sun up to about 0.06\,au between sheath density and the upstream solar wind density, the second (2a--2b) between sheath and ME density over 0.09--0.28\,au, and the third (3a--3b) covering 0.45--1.18\,au between ME density and upstream solar wind density. From that we may conclude, that at about 0.06\,au the sheath might become denser than the ambient solar wind, which could be interpreted as average starting distance for actual sheath formation. Between 0.09--0.28\,au, the sheath density overcomes the ME density. At this distance range the ME expansion may start to dominate the propagation phase, that leads i) to a strong decrease in density within the ME and ii) due to the relative enlargement over distance becomes an efficient piston-type driver causing plasma pile-up \citep{hundhausen72}. At distances starting around 0.45\,au and up to as far as 1.18\,au, the ME becomes lower in density than the ambient solar wind.

Complementary to Helios 1/2 data, we derived the average of ICME sheath and ME density measurements for a set of five PSP events observed at distances between 0.32--0.75\,au. The results support our findings from Helios data showing a similar trend between sheath and ME density as function of distance (cf., Figure~\ref{fig:15bs:ext}). However, we note that the PSP events are much weaker compared to Helios events. They do not drive a clear shock-sheath region, hence, do not give clear evidence for a strong pile-up or compression region close to the magnetic structure. Larger statistics from PSP (I)CME measurements closer to the Sun would be needed to derive more conclusive supporting evidence about sheath build up processes.  

\begin{table}[]
    \centering
    \begin{tabular}{|c|c|c|}
    \hline
    Dis. [au]  & Date  & Reference  \\ \hline
    0.32 & 20/Jan/2020 & \cite{joyce21} \\
    0.48 & 12/Sep/2020 & Bothmer \& Chifu list \\
    0.55 & 15/Mar/2019 & \cite{lario20} \\
    0.63 & 11/Feb/2021 & Bothmer \& Chifu list \\ 
    0.75 & 18/Sep/2021 & Bothmer \& Chifu list \\    \hline
    \end{tabular}
    \caption{PSP ICME event dates, distances (Dis.), and references.}
    \label{tab:psp}
\end{table}

For distances below 0.14\,au CME density results are available from indirect measurements analyzing coronagraph image data. Using multi-viewpoint SoHO and STEREO remote sensing white-light image data, the 3D CME mass and volume can be derived to calculate the ME density from \citep[][]{temmer21}. To compare with the extrapolated profiles from Helios in-situ measurements, we show in Figure~\ref{fig:15bs:ext} the results from the recent study by \cite{temmer21}, covering a sample of 29 CMEs observed during 2008--2014. The median ME density is derived from remote-sensing image data over the distance range 0.07--0.14\,au (i.e., SOHO/LASCO C3 coronagraph field of view). The results reveal an ME density that is substantially higher than the extrapolated ME density profile from Helios in-situ data. However, the lower ME density estimate (median minus standard deviation) is quite comparable. The discrepancies might be due to line-of-sight integration effects causing large uncertainties in the indirect method, namely 3D mass and volume derivation, as well as the different data set, i.e., solar cycle effects. We note that the steep drop in ME density during the early CME evolution might be a real effect that could be related to a different expansion behavior of CMEs close to the Sun. It is found that a strong lateral expansion dominates in the early CME evolution phase \citep[][]{patsourakos09,veronig18}, that could lead to a faster decay in ME density when close to the Sun.

Table~\ref{tab:relativ} summarizes for each structure over two distance bins, average values of density, plasma-beta, temperature, total magnetic field, speed and duration. For statistical reasons we divide the distance coverage 0.3--1\,au into two roughly similar distance bins with $r_1 <$ 0.7\,au and $r_2 >$ 0.7\,au, having sample sizes of 16 and 24 events, respectively. The density of the ME drops faster over distance than in the sheath and LE, which can be understood as a result of the strong ICME expansion. The plasma-beta of the sheath stays on average larger than 1, while the LE shows a clear increase over distance and the values for the ME are found to be similar. A possible interpretation for this behavior could be the steepening of the interface boundary because of increasing compression of the ME front similar to the heliospheric development of interfaces of co-rotating interaction regions as described, e.g., in \cite{forsyth99}. The LE decreases over distance in temperature and density as well as magnetic field, leading to an increase in the plasma-beta, which could be also a result of erosion \citep[][]{ruffenach12}. The sheath reveals slightly larger temperatures than the LE, and the speed of the LE is enhanced compared to the sheath, hence, showing that the sheath part might be driven by the LE. The magnetic field drop over distance is in the same range for all structures. The duration of the structures clearly increase with distance except for the LE. The average values for the ME are in accordance with the results given in \cite{bothmer98}.

\begin{table*}[]
    \centering
    \begin{tabular}{| c | c c c |}
\hline
 Parameter & Sheath & LE & ME \\
 \hline
 $N_p$ [cm$^{-3}$] & 108$\binom{243}{32}$ | 36$\binom{141}{9}$ & 105$\binom{265}{44}$ | 39$\binom{82}{8}$ & 58$\binom{105}{9}$ | 12$\binom{38}{3}$ \\ 
 $\beta$ & 1.1$\binom{4.6}{0.1}$ | 1.4$\binom{11.0}{0.2}$ & 0.8$\binom{1.8}{0.1}$ | 1.6$\binom{8.0}{0.1}$  & 0.1$\binom{0.3}{0.02}$ | 0.1$\binom{0.7}{0.04}$ \\  
 $T_p$ [$x10^4$ K] & 39$\binom{161}{6}$ | 26$\binom{120}{3}$ & 35$\binom{126}{8}$ | 19$\binom{88}{3}$ & 12$\binom{33}{4}$ | 11$\binom{70}{3}$   \\
 $B$ [nT] & 48$\binom{117}{16}$ | 17$\binom{47}{7}$ & 50$\binom{124}{15}$ | 16$\binom{43}{5}$ & 57$\binom{138}{28}$ | 19$\binom{43}{7}$  \\
 $v$ [km~s$^{-1}$] & 510$\binom{1100}{310}$ | 500$\binom{930}{350}$ & 540$\binom{1040}{300}$ | 510$\binom{980}{360}$ & 480$\binom{720}{350}$ | 480$\binom{730}{340}$ \\ 
 $t_{\rm dur}$ [h] & 2.7$\binom{6.6}{0.7}$ | 4.7$\binom{13.0}{0.5}$ & 2.5$\binom{10.5}{0.5}$ | 2.1$\binom{13.0}{0.6}$ & 17.7$\binom{35.5}{4.3}$ | 26.5$\binom{56.8}{1.6}$  \\
 \hline
    \end{tabular}
    \caption{Mean values together with their minimum and maximum range $\binom{\rm max}{\rm min}$ are given for two distance ranges $r_1$|$r_2$ with $r_1<0.7$\,au and $r_2>0.7$\,au. Parameters for the different structures given are, proton particle density ($N_p$), plasma-beta ($\beta$), temperature ($T_p$), total magnetic field strength ($B$), proton bulk speed ($v$), and duration ($t_{\rm dur}$). See also Figure~\ref{fig:2}, roughly depicting the average values given here.}
    \label{tab:relativ}
\end{table*} 

\begin{figure}
    \centering
    \includegraphics[width=0.85\hsize]{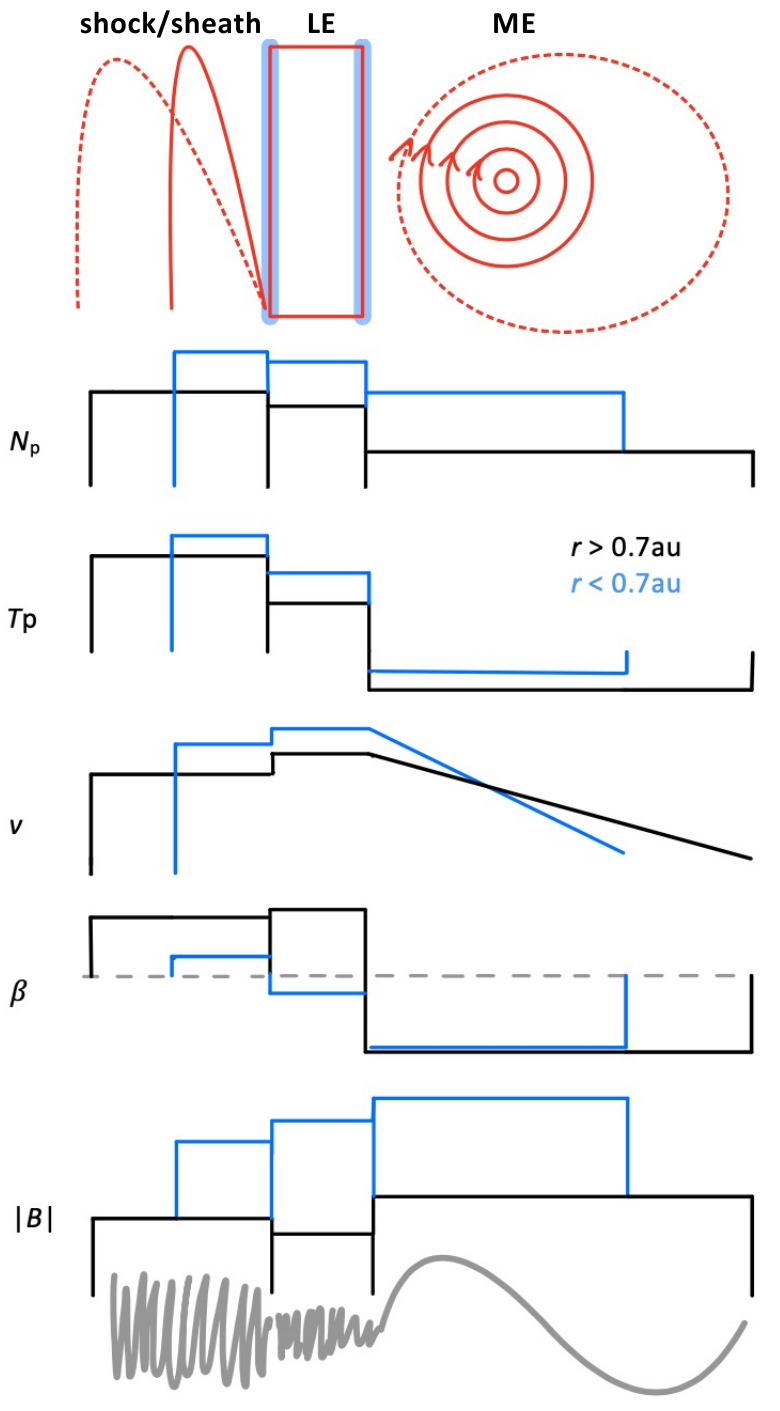}
    \caption{Cartoon illustrating the four different ICME structures and their relative parameter values. Top panel: discontinuities are marked as blue shaded areas before and after the LE, the expansion/evolution of the sheath and ME is marked with dotted red lines. Lower panels: particle density ($N_p$), temperature ($T_p$), speed ($v$), plasma-beta ($\beta$) and total magnetic field strength ($B$) with magnetic field variations from very fluctuating to smoothly rotating depicted below. For statistical reasons we derived the values for two different distance ranges and sketch the change in the average values accordingly ($r$<0.7\,au in blue and $r$>0.7\,au in black). For the actual average values and minimum/maximum ranges see Table~\ref{tab:relativ}. } 
    \label{fig:2}
\end{figure}

Figure~\ref{fig:2} gives a cartoon illustrating the four different ICME structures and their average parameter values. According to the statistical results given in Table~\ref{tab:relativ}, we show with two different colors the change of the parameter values over the two distance bins ($r_1 <$ 0.7\,au and $r_2 >$ 0.7\,au). This has implication for the comparison with remote sensing image data \citep[see also][]{vourlidas13}.

\section{Relating sheath and LE density to the ambient solar wind}\label{sec4}

The recent study by \cite{temmer21} showed another interesting result that is worth to investigate with the large statistical ICME sample from Helios data. At 1\,au a strong linear anti-correlation between the solar wind upstream speed (averaged over 24 hours before the arrival of the disturbance) and the measured sheath+LE density was derived ($cc=-0.73$). In comparison, the upstream solar wind density showed a less strong relation to the sheath density which is most likely due to stronger fluctuations of the density compared to the speed component of the solar wind \citep{gosling90,gosling91}. From that we may suggest that the ambient solar wind speed has some strong influence on the sheath build-up. Helios data could be used to empirically define relations between sheath density and solar wind speed as function of distance, which could be applied as additional input for ICME propagation models \citep[see also][]{takahashi17}. 

\begin{figure}
    \centering
    \includegraphics[width=\hsize]{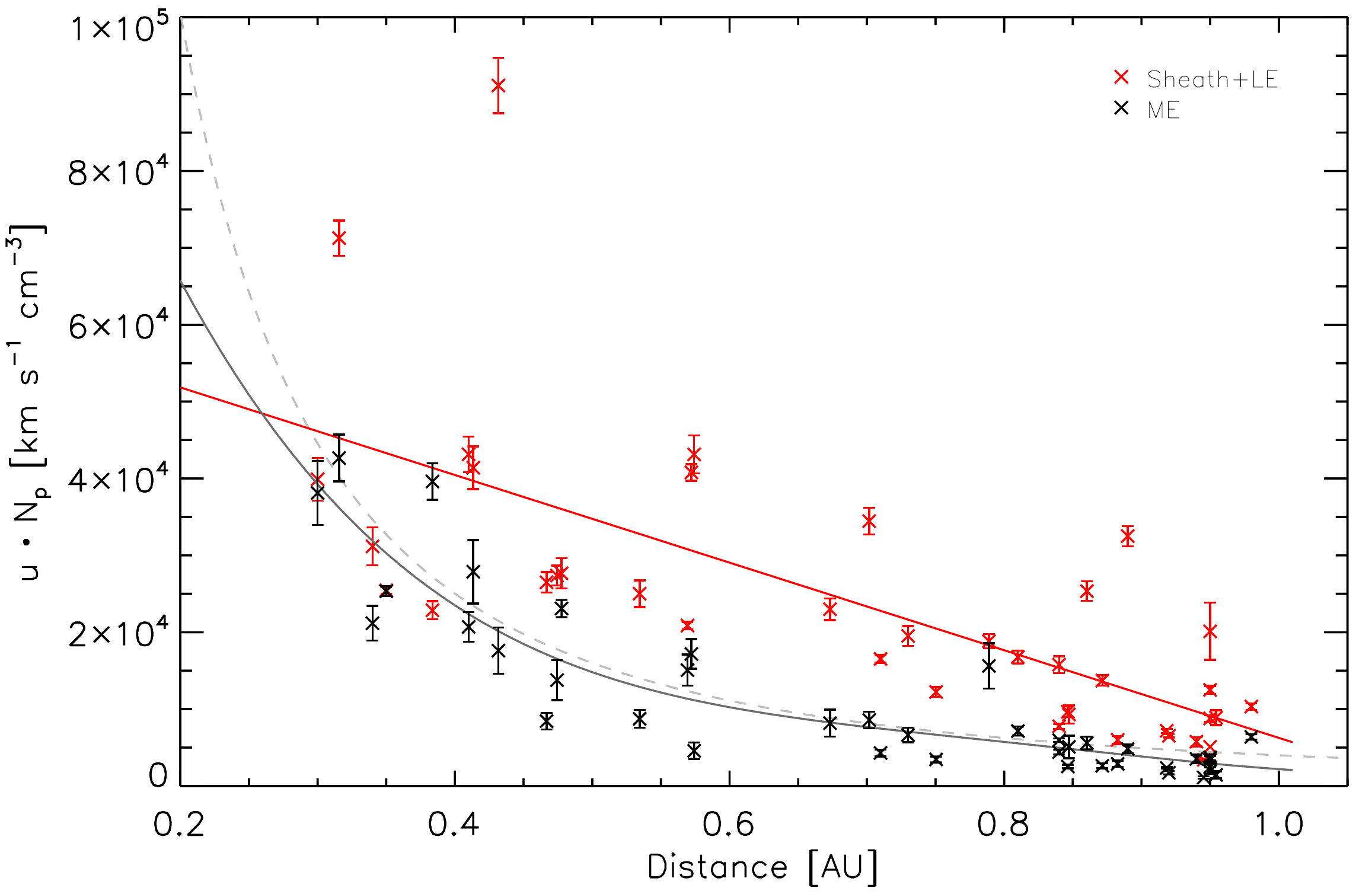} 
    \caption{Upstream speed, $u$, multiplied by the average density, $N_{\rm p}$, of the sheath+LE as well as the magnetic structure. The solid lines give the linear and fourth order polynomial fit, respectively. As comparison, the dashed line gives the empirical Leblanc solar wind density formula \citep[][]{leblanc98} for a 1~au density of 9~cm$^{-3}$, multiplied by the average upstream solar wind speed as derived for our data sample ($\sim$375~ km~s$^{-1}$). }
    \label{fig:empirical}
\end{figure}

As a first step we simply multiply the upstream solar wind speed $u$ (defined as three hours average before the arrival of the disturbance) with the average proton density $N_{\rm p}$ of the disturbance. Results for sheath+LE and ME structure are given in Figure~\ref{fig:empirical}. We find that the ME part can be fitted best with a 4$^{th}$ order polynomial fit, given by $$u N_p = 3.25\cdot10^5r^4-1.07\cdot10^6r^3+1.32\cdot10^6r^2-7.44\cdot10^5r+1.70\cdot10^5,$$ with the distance $r$ in au. At distances $r$>0.35\,au, the ME part becomes comparable to the solar wind speed-density profile derived from the Leblanc solar wind density model \citep{leblanc98} multiplied by the average upstream solar wind speed of $\sim$375~km~s$^{-1}$ as derived from our data sample (see dashed line in Fig.~\ref{fig:empirical}). On the other hand, the sheath+LE part is best fit with a linear function given by $$u N_p = -5.70\cdot10^4r+6.32\cdot10^4.$$ 
In a first approach, we can use these simple relations to estimate the density of sheath+LE and ME structure as function of distance and solar wind speed.

In a second approach, we separate the data into different distance bins and check the relation between solar wind speed and density of the identified ICME structures for each bin separately. As already defined in Sect.~\ref{sec3} we use $r_1$<0.7\,au covering 16 events and $r_2$>0.7\,au with 24 events. We derive for sheath+LE a mean density of $N_p(r_1)$=106.2~cm$^{-3}$ and $N_p(r_2)$=37.8~cm$^{-3}$ and for the ME $N_p(r_1)$=57.6~cm$^{-3}$ and $N_p(r_2)$=12.4~cm$^{-3}$. As comparison we also use four bins, being aware of the lower statistics, with $r_{\rm 1,2,3,4}$=[0.3--0.47, 0.47--0.75, 0.79--0.92, 0.92--0.98]\,au covering [9,11,11,9] events, from which we derive mean densities of $N_p(r_{1,2,3,4})$=[119.3, 78.4, 42.6, 22.4]\,cm$^{-3}$ for sheath+LE and $N_p(r_{1,2,3,4})$=[70.1, 32.4, 14.8, 7.6]\,cm$^{-3}$ for the ME. 

\begin{figure*}
    \centering
    \includegraphics[width=0.9\hsize]{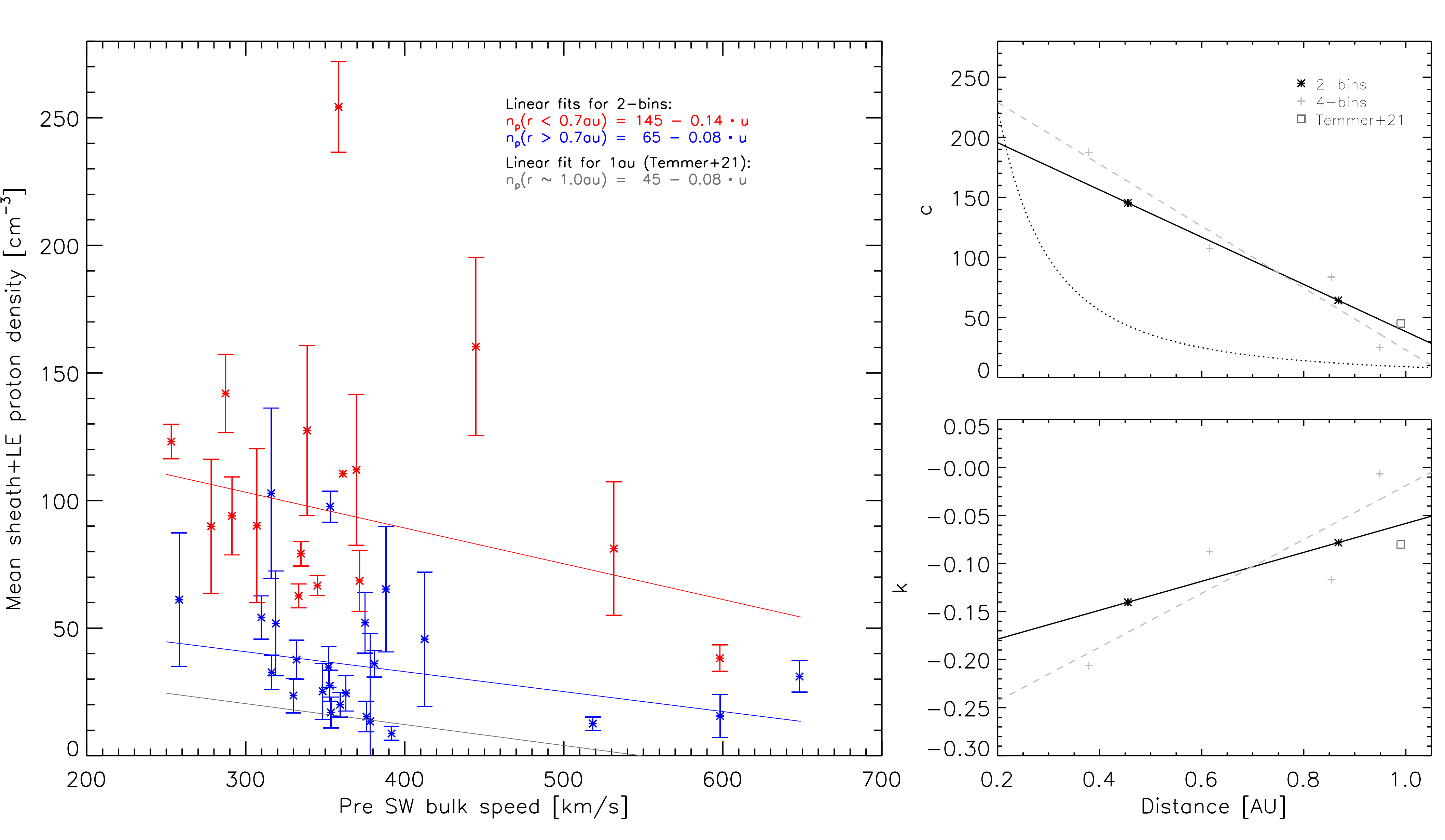}
    \caption{Left: Average density, $n_p$, and upstream solar wind speed, $u$, for two distance bins (red: $r_1$<0.7\,au covering 16 events; blue: $r_2$>0.70\,au covering 24 events). Right: Constants (top) and slope (bottom) for the fits from the 2-bin distribution, as given on the left, as well as for a 4-bin distribution ($r_{1,2,3,4}$=[0.3--0.47, 0.47--0.75, 0.79--0.92, 0.92--0.98]). In addition we give the results from \cite{temmer21} who derived for a sample of 29 events a linear relation between the upstream solar wind speed and sheath density for 1\,au. We show for each sample a linear fit and give the resulting regression formula in the legend.}
    \label{fig:emp0}
\end{figure*}

Figure~\ref{fig:emp0} (left panel) shows $N_p$ of sheath+LE structure in relation to $u$ for the entire data set and  color-coded for the two distance bins $r_1<$0.7\,au (red) and $r_2>$0.7\,au (blue). Not surprising, we derive a wide range of mean densities of 10--250\,cm$^{-3}$ that relate to a smaller range of upstream solar wind speed with 250--650\,km\,s$^{-1}$. When inspecting the distance bins, the density ranges become smaller and we may apply a simple linear fit to relate $u$ and $N_p$ in each bin which can be expressed by
$$N_p(u)=ku+c,$$
with $k$ the gradient and $c$ a constant. For comparison, we plot as gray line the density-speed relation from the 1\,au results as derived in \cite{temmer21}. The parameters for the linear fits are shown in the legend of the left panel of Fig.~\ref{fig:emp0}. When separating into 4 bins (not shown), the trend is the same, however, the scatter gets larger.

In a next step, we inspect the dependence of $c$ and $k$ over distance. Figure \ref{fig:emp0} (right panels) shows the derived constants and gradients from the fit for each distance bin as function of distance ($r$). In the top right panel values for $c$ are given derived from separating the sample into two and four bins, respectively. In addition, we show the constant from the linear relation from \cite{temmer21} for 1au. As comparison, we also plot the Leblanc solar wind density, normalized for $N_p$=9~cm$^{-3}$ at 1\,au (dotted line). The lower right panel shows $k$ values obtained from the linear fits using two bins, four bins, and results by \cite{temmer21}, respectively.

As can be seen, depending on the number of bins the constants $c$ and $k$ change slightly, but seem to follow a linear trend over distance, $r$ (given in au). We therefore may relate the parameters simply by 

$$c(r)=p_1(r)r+p_0$$ and
$$k(r)=q_1(r)r+q_0,$$

deriving $p_{0,1}$=[234.9,$-$196.8] and $q_{0,1}$=[$-$0.21,0.15] for the constant and gradient, respectively, from the linear fit using two bins. For four bins we derive $p_{0,1}$=[279.3,$-$255.1] and $q_{0,1}$=[$-$0.30,0.27]. From that we express an empirical relation to calculate the ICME sheath+LE density as function of distance and solar wind speed ahead of the ICME that can be given by

$$N_p(u,r)=k(r)u+c(r).$$

\begin{figure}
    \centering
    \includegraphics[width=\hsize]{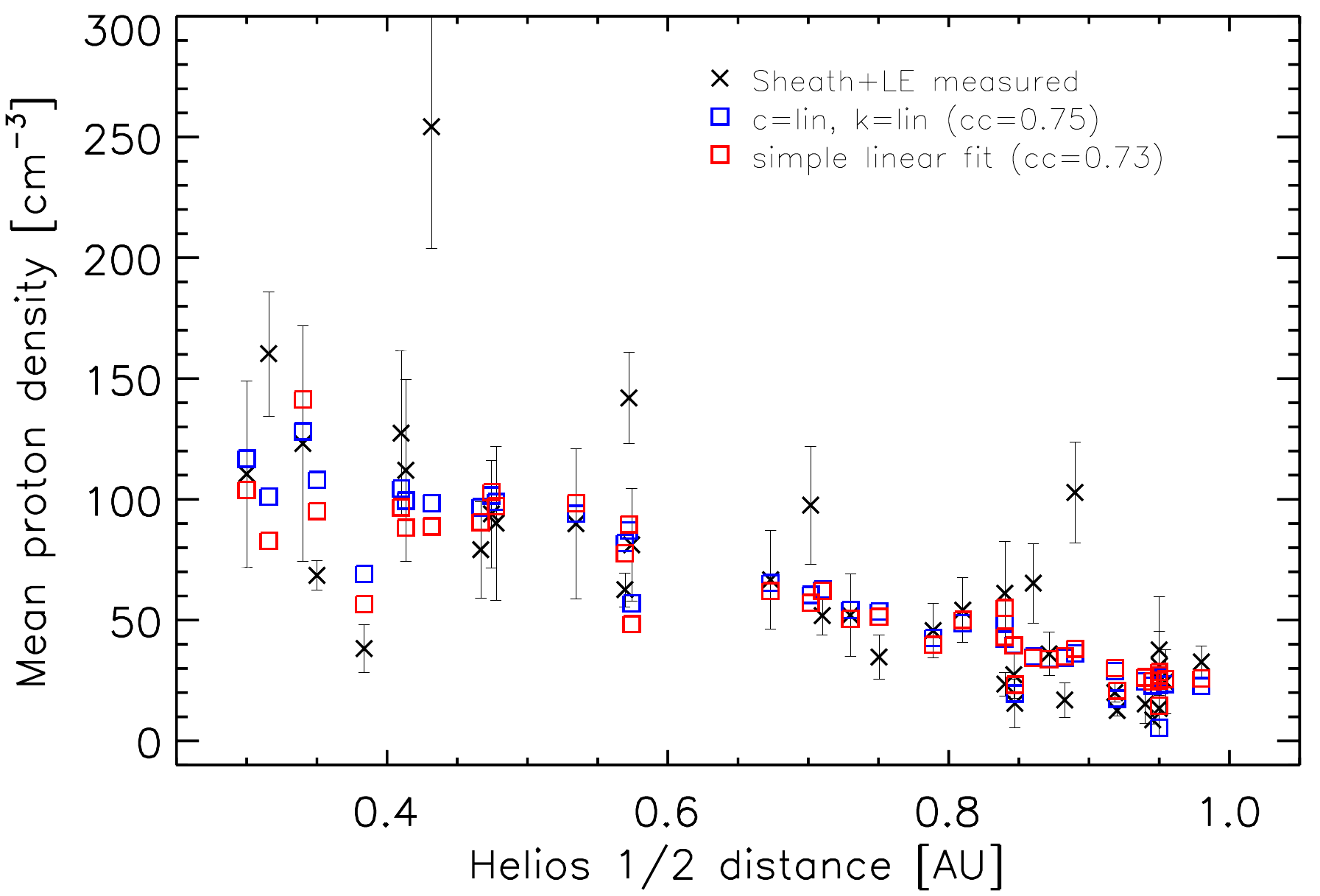}
    \caption{Measured average sheath and LE density including error bars, together with the results from the empirical relation between upstream solar wind speed and average density between shock and start of the magnetic structure (i.e., covering sheath and LE). Correlation coefficients between calculated and measured sheath+LE density are given as obtained from different empirical models. See legend for more details on the parameters.     }
    \label{fig:15}
\end{figure}

Figure~\ref{fig:15} shows the resulting sheath+LE densities as derived from the empirical relations described above in comparison to the measured sheath+LE densities over distance. The ``simple linear fit'' refers to the first approach using a linear fit to the sheath+LE density multiplied by the upstream solar wind speed. The ``c=lin, k=lin'' refers to the second approach when separating the upstream speed and density into two distance bins from which the linear fit parameters are calculated as function of distance. Applying these empirical methods we obtain a strong correlation with the measured sheath+LE densities with correlation coefficients of $cc$=0.73 for the simple linear fit and $cc$=0.75 when separating the data into distance bins. 

\section{Discussion and Conclusion}\label{sec5}

While the ICME magnetic ejecta (ME) region has been in the focus of most recent studies here we also investigate in detail the upstream regions ahead of the ME. Using a sample of 40 ICME events measured in-situ from Helios data and five from PSP, we were able to identify in most events four distinct density features, shock discontinuity, sheath region, leading edge (LE) and magnetic ejecta (ME). The LE of in-situ ICME events ahead of the magnetic ejecta body itself being of low plasma-beta, might represent naturally the bright leading edges seen in many coronagraph images of CME events. The in-situ data suggest that this enhancement is due to compressed ambient solar wind plasma and not part of the CME flux rope body as previously often thought of. The averaged density values of the structures, sheath, LE, and ME are investigated as function of distance from the Sun, together with other parameters such as speed, temperature, magnetic field, magnetic field variations, and estimated structure sizes.

We find that the density of the sheath decreases as linear function with $R^{-1.7}$ and that of the LE with $R^{-1.5}$. We also derive the results for ME density decrease as linear function with $R^{-2.4}$ and for the solar wind measured upstream of the disturbance with $R^{-2.1}$. The results for the ME density evolution are in accordance with previous findings from studies using Helios data as $R^{-2.4}$ \citep[][]{schwenn90,wang05}, $R^{-2.44}$ \citep[][]{leitner07}, and for the solar wind with $R^{-2.10}$ \citep[][]{schwenn90}, $R^{-2.01}$ \citep[][]{venzmer18}, or from modeling $R^{-2.07}$ \citep[][]{scolini21}. We use these results together with those for the sheath regions found from the current study, covering measurements in the range 0.3--1.0\,au, and extrapolate them to closer and slightly farther distances through which we obtain intersection regions. These show on the one hand, that the ME becomes lower in density than the ambient solar wind at a distance between 0.45--1.18\,au. In agreement, the study by \cite{wang05} found that the ME density becomes lower than the ambient solar wind at about 1.17\,au. Also \cite{liu05} report average ME densities that are slightly lower than the solar wind density at 1\,au. The lower density inside the ME in comparison to the ambient solar wind can be understood as a result of the ongoing ICME expansion \citep[see e.g., ][]{richardson06}. On the other hand, we find interesting results with respect to the density relation between sheath and ME as well as sheath and ambient solar wind for distances closer to the Sun. At about 0.06\,au the sheath density on a statistical basis, becomes higher than the ambient solar wind density. That result could be interpreted as the distance where the sheath starts to be formed. We note though, that we did a simple extrapolation from measurements derived at larger distances from the Sun. Physical processes might change below the sub-Alfvénic point which was recently measured by ground-breaking PSP measurements at a location of roughly 0.08\,au \citep[see][]{kasper21}. Nevertheless, from remote sensing image data it is obtained that 0.06\,au is about the distance that CMEs have fully developed their structure as observed in white-light data \citep[e.g.,][]{vourlidas00,pluta19}. At the latest at about 0.28\,au the ME density might fully drop below the sheath density which could be related to the stronger expansion of the ME in comparison to the sheath and/or a change in the expansion behavior of the ME. Recent model results by \cite{scolini21} report a more rapid ME expansion up to 0.4\,au compared to larger distances from Sun where the ME expansion is found to be more moderate. Interestingly, \cite{sachdeva17} found that on average at the distance of about 0.2\,au the drag force starts to dominate over the Lorentz force initially driving the CME. PSP results from the smaller sample of five events basically support the results derived with Helios data. However, we note that the PSP events are much weaker compared to Helios events, and do not drive strong shocks, hence, do not show the different structures so clearly. To confirm our results, future CME in-situ measurements from the SWEAP and FIELDS investigations together with remote sensing observations from WISPR \citep[][]{vourlidas16} aboard PSP for distances below 0.1\,au will be of great interest \citep[][]{bale16,kasper16}. 

From this study we find that the enhanced density structure ahead of the ME, i.e. the LE, is not the front of the flux rope itself, but rather a plasma region that consists of compressed or piled-up plasma, hence, has some width, but is not a structure that expands. The current study implies that the bright leading front of CMEs imaged by coronagraphs is similarly not the front of the CME flux rope (see Figure~\ref{fig:1a}). \cite{vourlidas13} interpreted the bright LE as pile-up of ambient coronal loops. We find that the size of sheath regions and ME structures increases with distance to the Sun, while the LE size is not showing significant variations. We are aware that in-situ we analyze localized variations and not all analyzed events reveal unique boundaries between individual features as is the case of the ICME observed by Helios 1 shown in Figure~\ref{fig:1}. The issues found are similar to the non-simultaneous appearances of typical ICME signatures, which were most often not observed all together \citep[][]{zurbuchen06}. It is also likely that the different identified regions depend on the CME/ICME kinematics with respect to the ambient solar wind and that they vary with distance from the Sun and depending on the crossing of the ICME in the different events. Assuming that the sheath is related to the shock-standoff distance, we used the relation between shock-standoff distance and obstacle/driver size as given by \cite{seiff62}, and compared these results to the observed sheath and LE size, respectively. For the sheath size a rather good match is found with a simple approximation of a spherical object. If one considers the LE as part of the sheath, as done in several other studies, the sheath size would increase in absolute size but without a radial trend. Comparing to previous studies, \cite{janvier19} derives for distances of 0.3/0.72/1\,au a sheath duration of 2.4/7.2./12~hours, respectively, and 7.2/14.4/19.2~hours for the ME duration. \cite{kilpua13} obtain for 1\,au an average sheath duration of 9.1$\pm$4.0~hours, 7.1$\pm$4.1~hours for the front region and 20.6$\pm$9.1~hours for the ME. The current study shows for the sheath size a less strong increase over distance. Even summing up sheath and LE regions, we derive 5.2~hours (r$<$0.7au) and 6.8~hours (r$>$0.7au) which is much shorter in comparison. The ME is found to be larger on average with 17.7 and 26.5~hours, respectively, for r$<$0.7 au and r$>$0.7au. We note that the Helios data set covers a different solar cycle in comparison to the studies from \cite{kilpua13} and \cite{janvier19}, hence, results might deviate from the more recent events. We also note that the different ICMEs might have been intersected at different distances from the apex leading to deviations in the sheath and leading edge region sizes and properties as already pointed out. 

For the sheath size we find the highest correlation with the ambient solar wind density ($cc=-$0.52) and the plasma-beta of the ME ($cc=-$0.51). A weak correlation between the sheath size and driver speed (LE, ME) exists, as well as a weak anti-correlation to the sheath density itself. \cite{masias16} found from ICME measurements at 1\,au, that slow ejecta on average drive a more massive sheath region. In total, the statistics in the Appendix also shows, that the ME magnetic field relates very strongly to the LE and sheath density. With that we support the results by \cite{salman21} who concluded that ME properties shape the sheath properties and that slower CMEs, spending longer time in the solar wind, do not drive larger sheaths. These results are important for better understanding the CME mass evolution due to sheath enlargement. We also support results from \cite{maricic20} who derived strong correlations between the magnetic field in the sheath and leading edge, indicating a strong physical relation between these two structures.

The study by \cite{temmer21} found for a set of 29 ICME data at L1, a strong anti-correlation ($cc=-$0.73) between sheath+LE density and the ambient solar wind speed measured 24 hours ahead of the disturbance. In contrast, the current study covers a distance range from 0.3\,au to 1\,au. For that we find the sheath density to be moderately anti-correlated with the local solar wind plasma speed upstream of the ICME shock ($cc=-$0.41). Repeating the analysis of the upstream solar wind speed-sheath density relation for two different distance bins ($r_1$, $r_2$), we derive for the events measured at distances larger than 0.7\,au an increase in the Spearman correlation coefficient with $cc=-$0.58 (80\% confidence level). For events closer to the Sun the relation is less clear and the scatter increases. Assuming there is a local dependency between the sheath density and the ambient solar wind speed, these results allow to define some empirical relations. Those enable to model the density of sheath and LE structure over distance simply by knowing the upstream solar wind speed at that distance. This supports the more reliable model output of the background solar wind speed component and the results presented here could be implemented in numerical models in order to add the pile-up or compression of material ahead of the propagating ICME. In a similar way \cite{kay20} relates the downstream density and magnetic field strength to the upstream properties from which an empirical model is derived to predict the sheath structure at 1\,au. 

\vspace{0.2cm}
We summarize our main findings:
\begin{itemize}
    \item Four main density structures, namely shock, sheath, leading edge, and magnetic ejecta are identified in plasma and magnetic measurements of 45 studied ICMEs observed by the Helios 1 and 2 and PSP spacecraft.
    \item The radial distance from the Sun at which the sheath forms is estimated to about 0.06\,au where the inferred sheath density exceeds the ambient solar wind density.
    \item The sheath density might start to become prominent over the ME density in the distance range 0.09--0.28\,au where the ME expansion should likely be stronger than further out in the heliosphere.
    \item The ME density is inferred to become lower than the ambient solar wind density in the distance range from 0.45--1.18\,au.
    \item The sheath characteristics seem to be related to the upstream solar wind and ME properties.
    \item Assuming a local linear relation between sheath density and ambient solar wind speed, we give empirical relations that could support CME propagation models.
\end{itemize}

Investigating the shock-sheath and leading edge structures separately as function over distance using Helios data, is an important first step and shows interesting results where these structures might form. With PSP approaching in near future the Sun as close as 0.05\,au, we will certainly detect more and stronger CME events to obtain measurements that might re-affirm the presented results and give more detailed insight about the sheath build up processes. With increasing solar activity we expect also stronger events to get more conclusive results \citep[see also][]{venzmer18}. Moreover, the Wide-field Imager for PSP \citep[WISPR][]{vourlidas16} enables comparative studies with remote sensing white-light data with high spatial resolution.

\begin{acknowledgements} We thank the anonymous referee for the careful reading and the very constructive comments that helped to improve the manuscript. VB acknowledges the support of the Coronagraphic German And US Solar Probe Survey (CGAUSS) project for WISPR by the German Aerospace Centre (DLR) under grant 50OL1901 as a national contribution to the NASA Parker Solar Probe mission. We thank Iulia Chifu for her support in PSP data provision. Parker Solar Probe was designed, built, and is now operated by the Johns Hopkins Applied Physics Laboratory as part of NASA’s Living with a Star (LWS) program (contract NNN06AA01C). Support from the LWS management and technical team has played a critical role in the success of the Parker Solar Probe mission. For the access to the PSP FIELDS and SWEAP data we acknowledge the Parker Solar Probe Science Gateway at \url{https://sppgway.jhuapl.edu}.
\end{acknowledgements}

\bibliographystyle{aa}

\begin{appendix}
\section{Spearman correlation coefficient matrix for different solar wind parameters}

\begin{figure*}
    \centering
    \includegraphics[width=\textwidth]{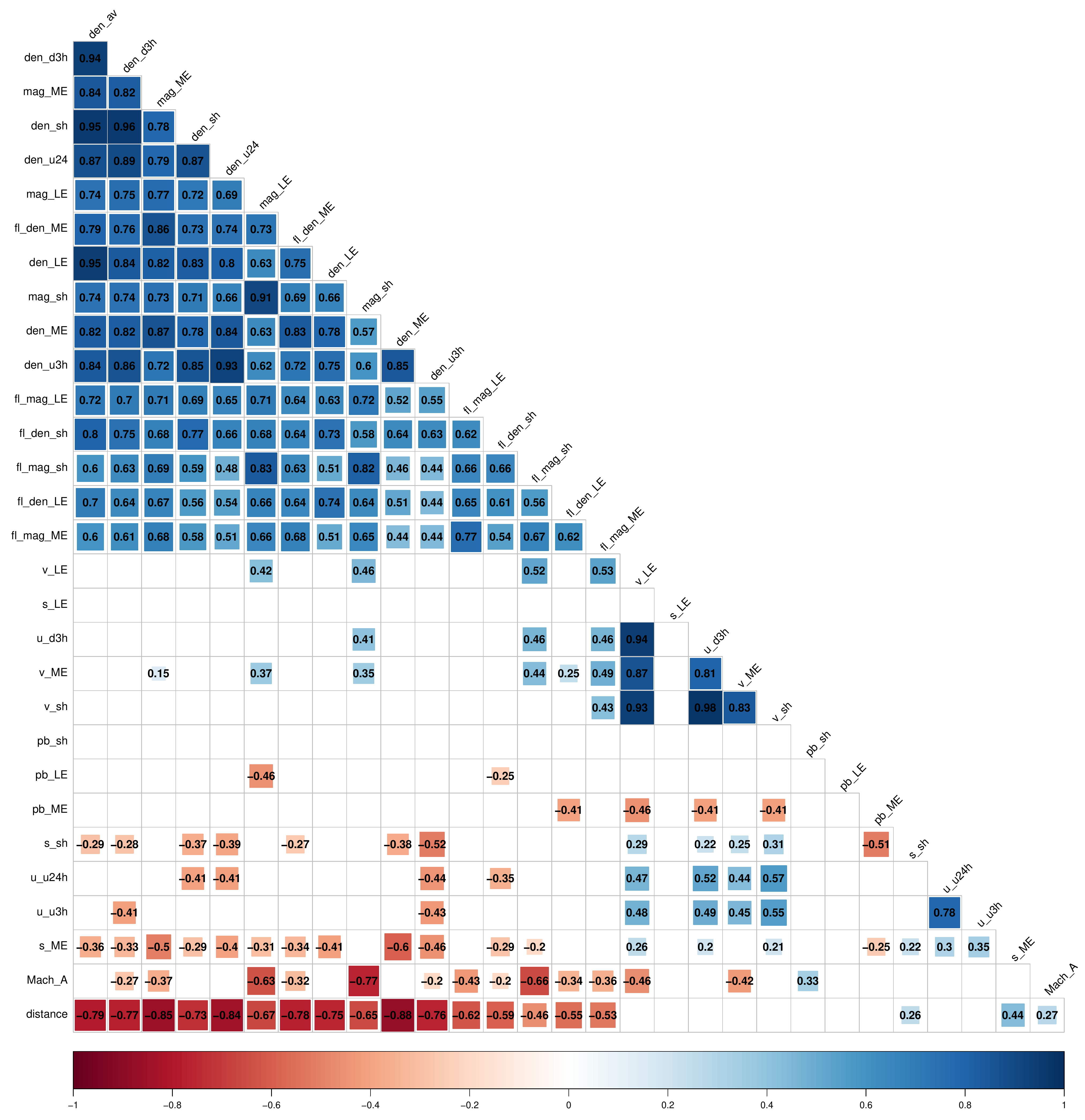}
    \caption{The Spearman correlation coefficients are calculated on a 90\% significance level. Correlations with significance below are not given (i.e., left blank).}
    \label{fig:rplot}
\end{figure*}

Figure~\ref{fig:rplot} shows the statistical relation between various parameters extracted from Helios 1/2 plasma and magnetic field data. The Spearman correlation coefficients are calculated on a 90\% significance level. Correlations with significance below are not given (i.e., left blank). The matrix covers for each structure sheath (sh), leading edge (LE) and magnetic ejecta (ME) their average parameters in density (den), magnetic field (mag), speed (v), and respective fluctuations (fl) from the standard deviation of these parameters. Furthermore we give the plasma-beta (pb), upstream solar wind speed measured 3 and 24 hours, respectively, ahead of the disturbance (u\_u3h, u\_u24h), as well as downstream solar wind density and speed, respectively, measured 3 hours after the disturbance (d\_d3h, u\_d3h). The calculated correlations additionally cover the size (s) of each structure, the Alfvén Mach number (Mach\_A, see Section~\ref{sec3}), and the distance at which the measurements were done.

\begin{figure*}
    \centering
    \includegraphics[width=1.\hsize]{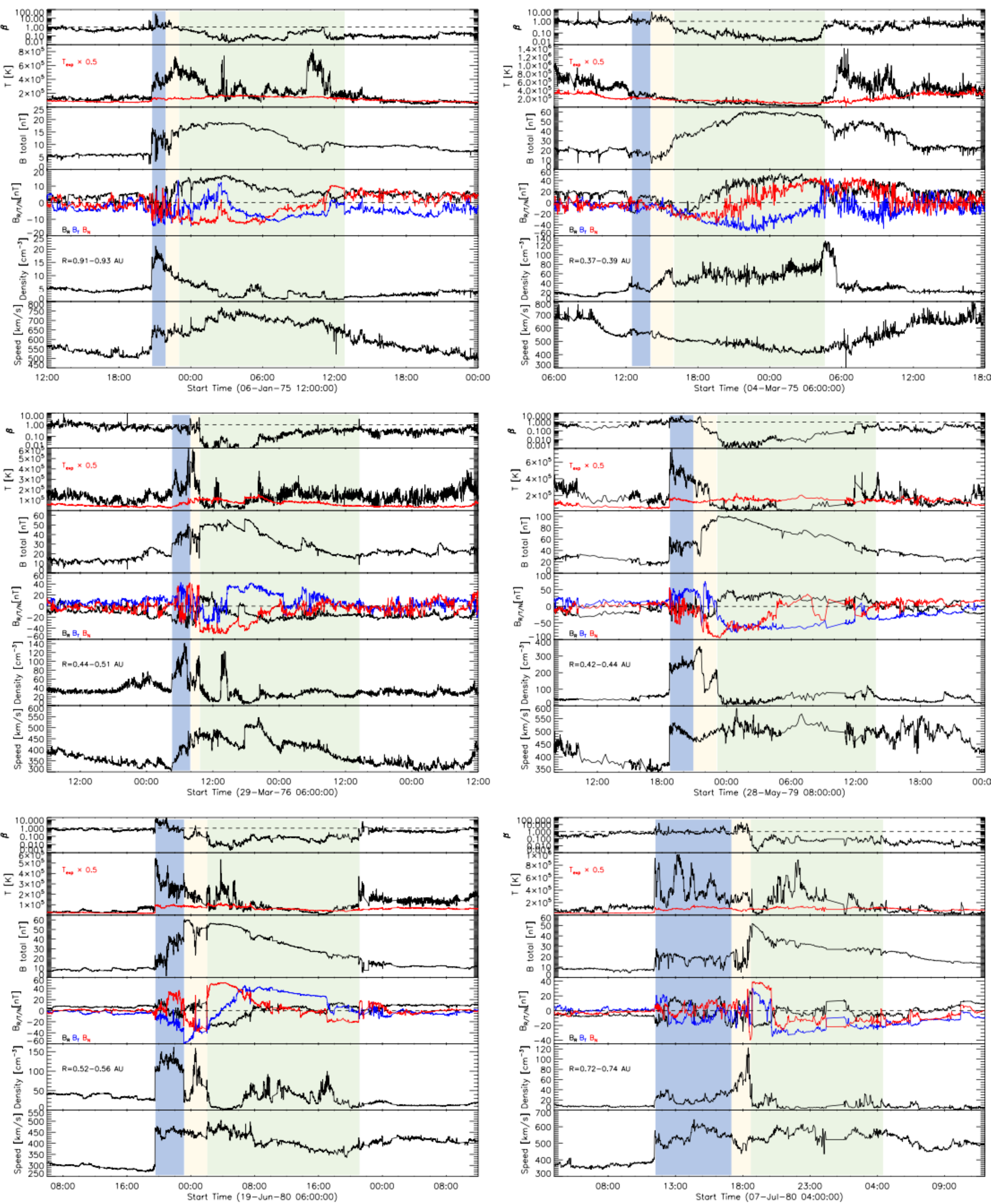}
    \caption{Sample ICME events observed with Helios 1 and 2 at different distances from the Sun. See Table~\ref{tab:helios} for details.}
    \label{fig:helios_app}
\end{figure*}

\begin{figure*}
    \centering
    \includegraphics[width=1.\hsize]{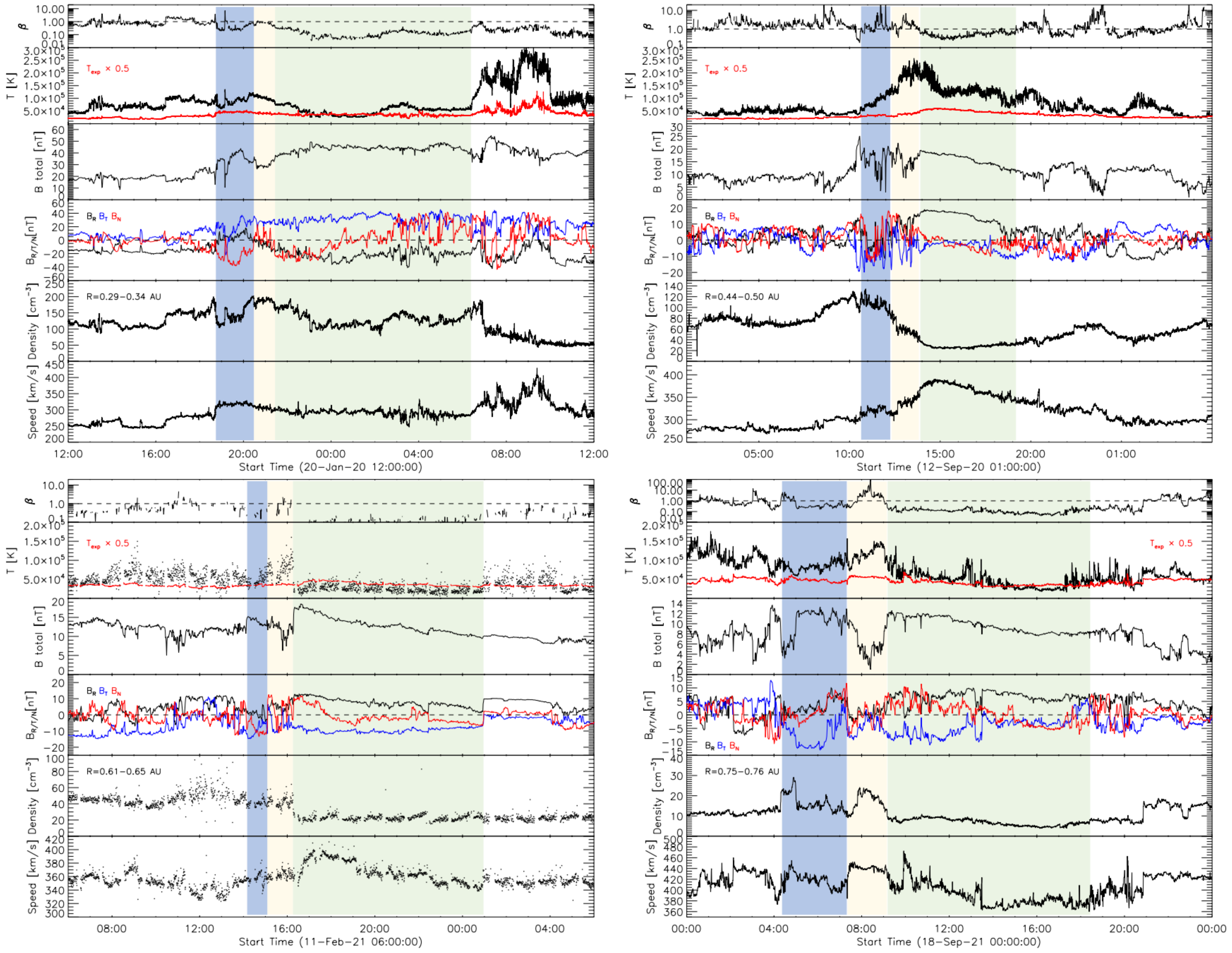}
    \caption{ICME samples observed with PSP. See Table~\ref{tab:psp} for details.}
    \label{fig:psp_app}
\end{figure*}

\begin{table*}[]
    \centering
\begin{tabular}{ccccc|ccccc}
s/c & year & DOY    & R {[}au{]} & list & s/c & year & DOY     & R{[}au{]} & list \\ \hline
H1  & 1975 & 7       & 0.92       & B\&S & H1  & 1980 & 90      & 0.88      & B\&S \\
H1  & 1975 & 63,64   & 0.39       & B\&S & H1  & 1980 & 162,163 & 0.41       & B\&S \\
H1  & 1975 & 92      & 0.48       & B\&S & H1  & 1980 & 171,172 & 0.53      & B\&S \\
H1  & 1975 & 313     & 0.87       & B\&S & H1  & 1980 & 175     & 0.57      & B\&S \\
H1  & 1976 & 187     & 0.98       & B\&S & H1  & 1980 & 189     & 0.73      & new  \\
H1  & 1977 & 29,30   & 0.95       & B\&S & H1  & 1980 & 203     & 0.84      & new  \\
H1  & 1977 & 78,79   & 0.57       & B\&S & H1  & 1981 & 103     & 0.89      & new  \\
H1  & 1977 & 159,160 & 0.86       & B\&S & H1  & 1981 & 117,118 & 0.79      & B\&S \\
H1  & 1977 & 240,241 & 0.84       & B\&S & H1  & 1981 & 146,147 & 0.48      & B\&S \\
H1  & 1977 & 268,269 & 0.57       & B\&S & H1  & 1981 & 156     & 0.35      & new  \\
H1  & 1977 & 335     & 0.75       & B\&S & H1  & 1981 & 170     & 0.34      & B\&S \\
H1  & 1978 & 3,4     & 0.95       & B\&S & H2  & 1976 & 90      & 0.47      & B\&S \\
H1  & 1978 & 46,47   & 0.95       & B\&S & H2  & 1977 & 76      & 0.71      & B\&S \\
H1  & 1978 & 61,62   & 0.87       & B\&S & H2  & 1977 & 335     & 0.70      & new  \\
H1  & 1978 & 132     & 0.41       & new  & H2  & 1978 & 4,5     & 0.94      & B\&S \\
H1  & 1978 & 363,364 & 0.85       & B\&S & H2  & 1978 & 47,48   & 0.95      & B\&S \\
H1  & 1979 & 58,59   & 0.96       & B\&S & H2  & 1978 & 114     & 0.32      & B\&S \\
H1  & 1979 & 62      & 0.94       & B\&S & H2  & 1978 & 358,359 & 0.85      & B\&S \\
H1  & 1979 & 148,149 & 0.43       & B\&S & H2  & 1979 & 93      & 0.68      & B\&S    \\
H1  & 1980 & 82,83   & 0.92       & B\&S & H2  & 1979 & 129     & 0.30      & B\&S \\
\end{tabular}
 \caption{List of ICME Helios events with year and day of year (DOY), spacecraft (s/c) providing the in-situ measurements, Helios 1 (H1) or Helios 2 (H2), and s/c distance from Sun (R). 34 events are covered in the catalogue from B\&S, 6 new ones were added in the frame of this study.}
    \label{tab:helios}
    \end{table*}
   
\end{appendix}
\end{document}